# Next-Level, Robotic Telescope-Based Observing Experiences to Boost STEM Enrollments and Majors on a National Scale: Year 1 Report

Daniel E. Reichart, Joshua Haislip, Vladimir Kouprianov, Ruide Fu, Logan Selph, Shengjie Xu

*University of North Carolina at Chapel Hill, Department of Physics and Astronomy, Campus Box 3255, Chapel Hill, NC 27599-3255*

John Torian, Jonathan Keohane

*Hampden-Sydney College, Department of Physics and Astronomy, Hampden-Sydney, VA 23943*

Daryl Janzen

*University of Saskatchewan, Department of Physics and Engineering Physics, 116 Science Place, Saskatoon, SK, Canada S7N 5E2*

David Moffett

*Department of Physics, Furman University, Greenville, SC 29613*

Stanley Converse

*Wake Technical Community College, Department of Mathematics and Physics, 6600 Louisburg Road, Raleigh, NC 27616*

**Abstract**. Funded by a $3M Department of Defense (DoD) National Defense Education Program (NDEP) award, we are developing and deploying on a national scale a follow-up curriculum to "Our Place In Space!", or OPIS!, in which ≈3,500 survey-level astronomy students are using our global network of "Skynet" robotic telescopes each year. The goal of this new curriculum, called "Astrophotography of the Multi-Wavelength Universe!", or MWU!, is to boost the number of these students who choose STEM majors. During Y1, our participating educators have developed MWU!'s (now renumbered) 2nd and 4th modules, and are in the process of developing its 3rd and 7th modules (out of 7). Solid progress has also been made on the software front, (1) where we have developed new graphing/analysis/modeling interfaces in support of Modules 2 and 4, and in response to feedback from the participating educators; and (2) where we are in the process of developing and adding astrophotography capabilities to Afterglow Access (AgA), our student-level, web-based, image processing and analysis application, in support of Modules 1 – 3 and 5 – 7. On the hardware front, development of our first four signal-processing units proceeds on schedule; these are key to Skynet's integration of a global network of radio telescopes, capable of exploring the invisible universe. Preparations have also been made on the evaluation and accessibility fronts, for when the first MWU! modules are deployed in Spring 2023.

## 1.  Programmatic Overview

Over the past two decades, UNC-Chapel Hill has built one of the two largest networks of fully automated, or robotic, telescopes in the world, significantly advancing this new technology. These telescopes are used both for cutting-edge research and for cutting-edge education: Funded by a series of large NSF awards, (1) we have developed unique, student-level, observing and image-analysis interfaces, allowing students, of all ages, to use this globally distributed, research tool, right alongside the professionals; and (2) often in partnerships with professional educators and education researchers, we have developed a sequence of observation-based curricula and experiences that leverage these hardware and software resources, from the elementary-school level through the graduate-school level, reinforcing and strengthening the STEM pipeline. Now funded by DoD NDEP, we are developing a new, primarily undergraduate-level curriculum that will target students as they are deciding their major/minor. This curriculum, called "The Multi-Wavelength Universe!", or MWU!, will leverage two of our most successful education efforts to date [1,2], with the end goal of significantly boosting STEM enrollments and STEM majors and minors on a national scale, as well as boosting students' technical and research skills.

## 2.  Curriculum

Our participant program began on 9/30/21, 15 days after the start of the award. Currently 26 educators are participating; most are faculty members at the participating institutions. The educators are divided into two groups, and each group is meeting weekly.

Each group is responsible for producing 3 – 4 of the 7 modules of the "Astrophotography of the Multi-Wavelength Universe!", or MWU!, curriculum, and trialing them with their students. We are currently trialing the 2nd and 4th modules of the curriculum, and are actively developing the $3^{rd}$ and $7^{th}$ modules:

### 2.1.  Module 2

In this module, students break into teams and each team uses Skynet to acquire images of one or more star clusters, each at three or more visible-light wavelengths. From these images, they measure the brightness of every detectable star, and calibrate these values. They then use these data to plot HR diagrams, and fit isochrone models to these data to date the cluster (see Figures 1 and 2). Lastly, they combine their multi-wavelength images into a single, deep color picture (see Figure 3), and develop an intuitive understanding of the relationship between ages and colors of stellar populations. The current version of Module 2 can be found here:

https://openpress.usask.ca/skynet/chapter/hr-diagrams/

### 2.2.  Module 3

In this module, students will break into teams and each team will use Skynet to produce two deep (high signal-to-noise), color pictures of star birth and star death, respectively, using red, green, and blue filters as in Module 2, but also using broadband and narrowband filters. For star birth, they will image a star-forming region, focusing on its emission, reflection, and/or dark nebulae (see Figure 4). For star death, they will image either a "planetary" nebula (see Figure 5) or a supernova remnant, focusing on narrowband emission from hydrogen, oxygen, and/or sulfur. Students will also learn how to supplement their images with archival data, primarily at infrared wavelengths, e.g., acquired with space-borne, NASA observatories.

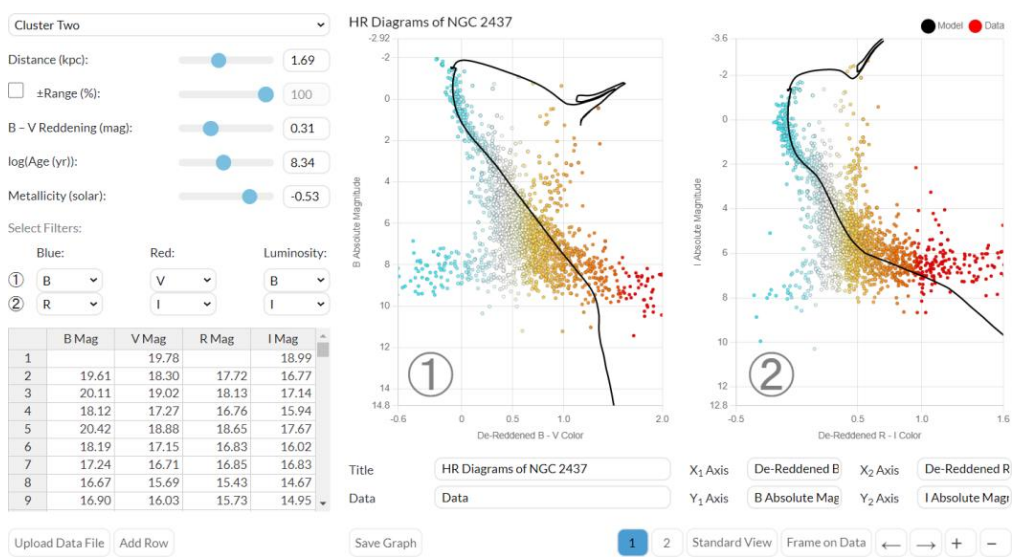

Figure 1. HR diagrams of Messier 46, with stellar brightnesses and colors measured from Figure 3 using Afterglow Access (AgA; see Software, below) and analyzed with our "Cluster Two" interface. Isochrone modeling can be challenging because of the great many unrelated, field stars in the image.

This group has been focusing on the astrophotography component of this module, trialing these activities and developing "how-to" guides. However, they are now turning to developing the more quantitative, physical components of this module. The goal is for them to complete this work in the spring, producing a manual similar to those of Modules 2 and 4.

**2.3. Module 4**

In this module, students use Skynet to observe a variety of bright, slow pulsars at radio wavelengths, to introduce them to non-thermal continuum-emitting mechanisms. Some of the pulsars are sufficiently bright to see individual pulses; others are lost in the noise and extracted by identifying the pulsar's period, and then folding the data at this period, to beat down the noise. By calibrating and differencing orthogonal polarization channels in this period-folded data, students show the emission to be polar-ized, and hence non-thermal (see Figures 6 – 8). The current version of Module 4 can be found here:

https://openpress.usask.ca/skynet/chapter/radio-observations-of-pulsars/

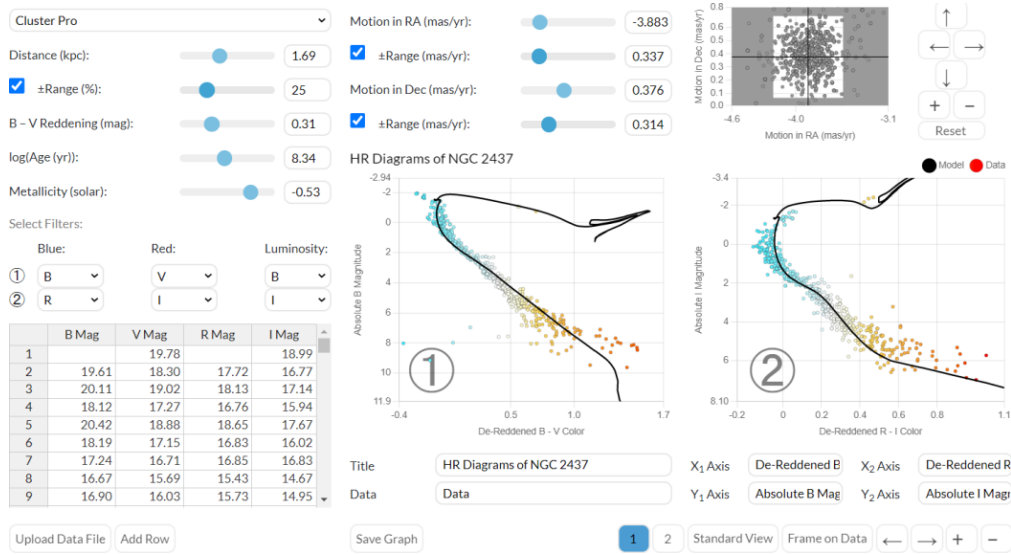

Figure 2. The same data as in Figure 1, but with our new "Cluster Pro" interface, students can now quickly isolate and view only the cluster stars, based on their common distance away and their common proper motion. This information is now added automatically upon file upload by cross-matching with astronomy's Gaia database. Isochrone modeling then cleanly reveals this star cluster to be ≈250 million years old.

### 2.4. Module 7

In this module, students will break into teams and each team will use Skynet to produce deep, color pictures of at least two galaxies, using red, green, and blue filters as in Module 2, but also using broadband and narrowband filters, and archival infrared data, as in Module 3. For one, they will image a spiral and/or starburst galaxy, where the light is dominated by younger, bluer stars (Module 2), pink star-forming regions (SFRs, Module 3), and dark, dusty clouds (Module 3; see Figure 9), as well as by line emission from cold, hydrogen gas in radio spectra. For the other, they will image an active, elliptical galaxy. Such galaxies are dominated by older, redder stars (Module 2), and are devoid of SFRs, dust, and cold, hydrogen gas. However, active ellipticals are in the process of interacting with/absorbing another galaxy, which feeds its central, massive black hole, and produces radio-emitting jets (which they will also image with Skynet's radio telescopes, re-applying what they will learn in Module 5). Such objects demonstrate how galaxies evolve, and are sufficiently rich in content that they will serve as MWU's capstone project, consolidating skills from multiple MWU! modules.

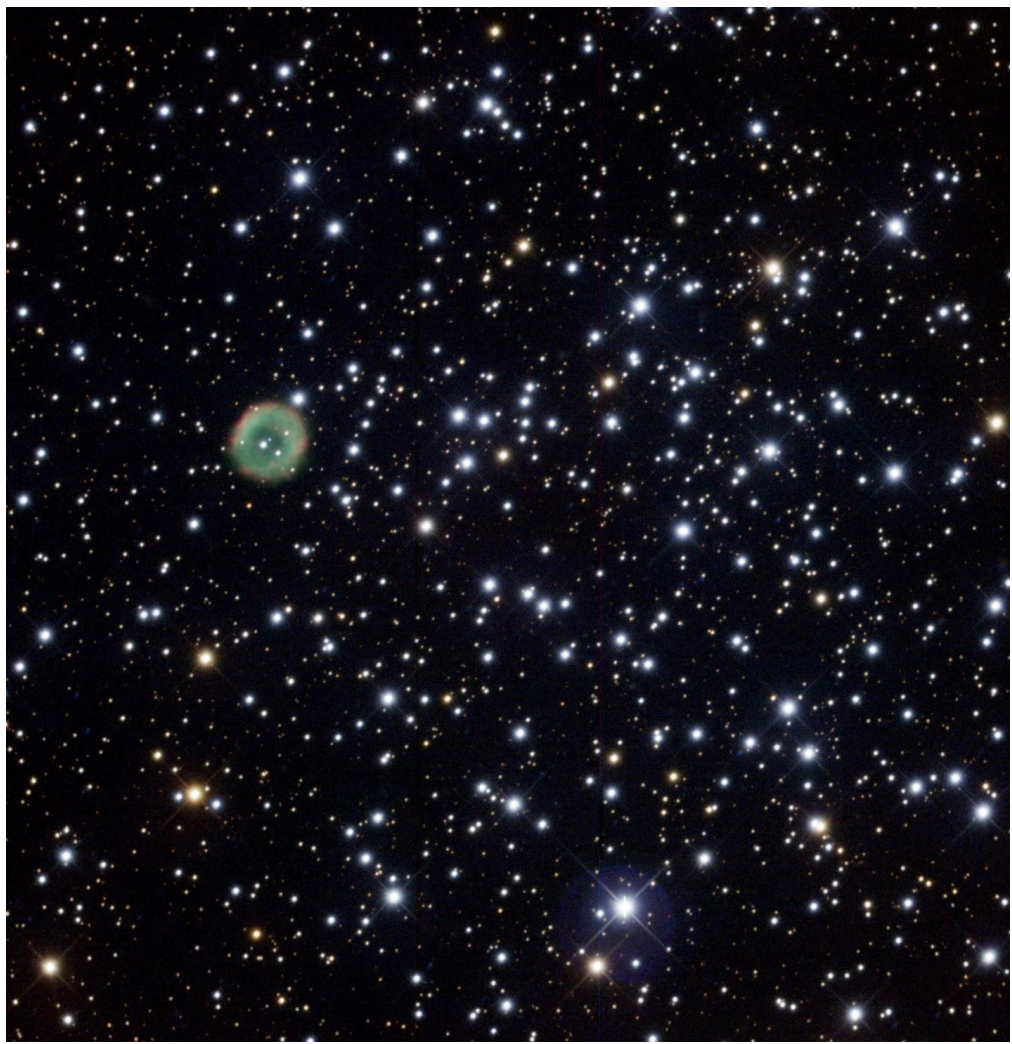

Figure 3. RGB combination of Messier 46, using Skynet's PROMPT-6 telescope at Cerro Tololo Inter-American Observatory (CTIO). All processing, including the astrophotography component, was done using AgA. In Module 2, students learn: (1) how to align and stack images, removing bad pixels and cosmic rays; (2) RGB combination, calibrating each layer using astronomical catalogs to achieve natural color, at least as viewed from Earth; (3) how to correct each layer for reddening by intervening dust, to recover the intrinsic natural color of the target; and (4) how to "stretch" an image, to bring out details at both its faint and bright ends simultaneously.

This group has been focusing on the astrophotography component of this module, trialing these activities and developing "how-to" guides. However, they are now turning to developing the more quantitative, physical components of this module. The goal is for them to complete this work in the spring, producing a manual similar to those of Modules 2 and 4.

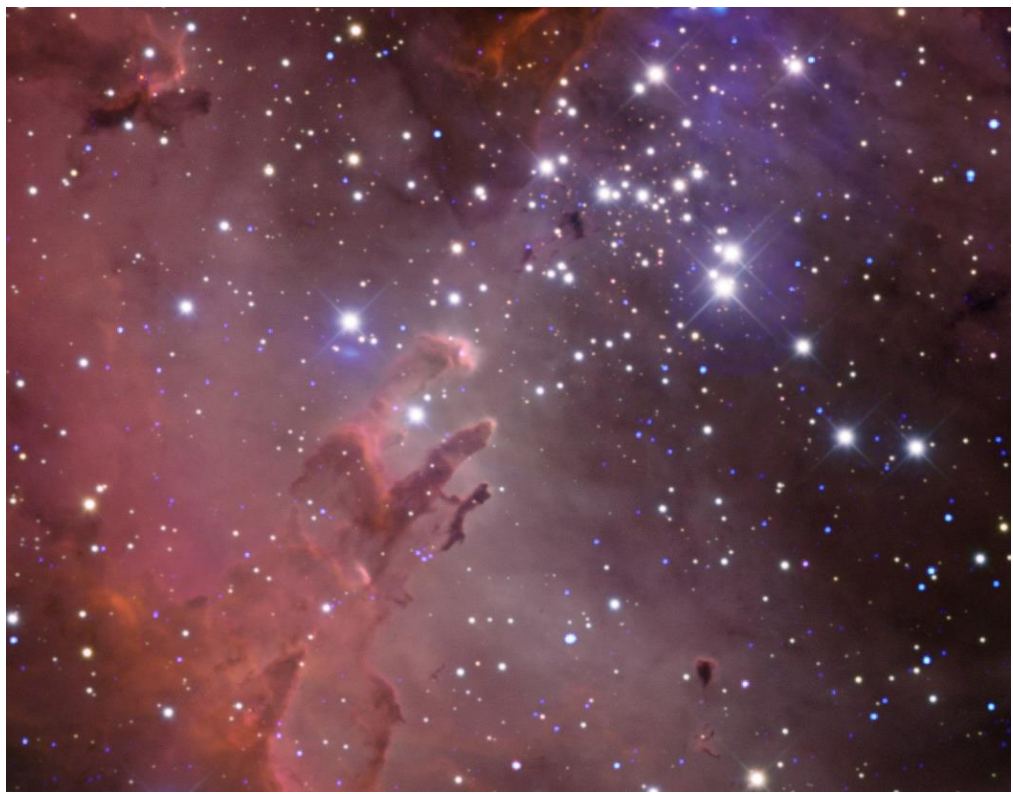

Figure 4. LRGB+NB+IR combination of Messier 16, using Skynet's PROMPT-6 and PROMPT-7 telescopes at CTIO, and supplemented with archival infrared data. All processing was done using AgA. Here, young, hot stars (upper right) are ionizing the surrounding gas, which then recombines, emitting Hα (pink), OIII (blue-green), and SII (red) line emission. These stars also heat the surrounding, dusty clouds that this hot gas is pushing against, causing this dust to re-radiate at infrared wavelengths, added in here as orange (8 micron) and violet (24 micron), from NASA's Spitzer spacecraft. Lastly, this dust also hides stars at visible wavelengths, but not at infrared wavelengths, depicted here as blue (2.2 micron), from the ground-based 2MASS survey. Students learn: (1) how to use a broadband luminance layer, to either boost an image's broadband, stellar content, or diminish it in favor of narrowband, gas content; (2) how to add in narrowband layers, to emphasize gas content; and (3) how to find and add in archival, infrared layers, to emphasize dust content, as well as dust-hidden stars. Students will also use information in their image to estimate the temperature and density of the gas, both in the ionization region, and in the surrounding, dusty cloud.

## 3.    Software

The most immediate software need in Y1 was developing the graphing/analysis/modeling interfaces that the educators needed to develop Modules 2 and 4, and that are now being trialed by their students (Figures 1, 2, 6, 7, and 8). We decided to add these to the collection of interfaces that we have already developed for "Our Place In Space!", or OPIS!, the precursor curriculum for MWU!: https://tinyurl.com/skynet-graph.

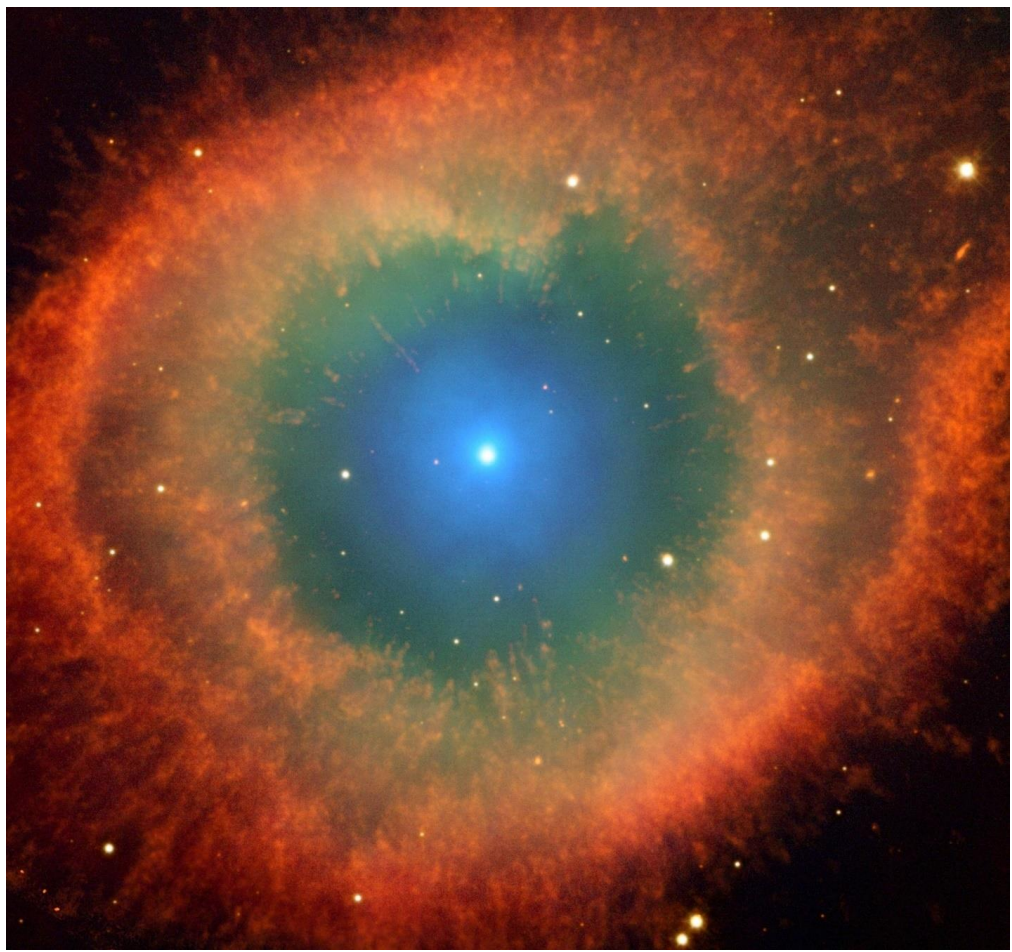

Figure 5. LRGB+NB+IR combination of planetary nebula NGC 7293, using Skynet's PROMPT-6 telescope at CTIO, and supplemented with archival infrared data. All processing was done using AgA. Here, stellar colors are natural, but diminished using the luminance layer in favor of the image's narrowband Hα (red) and OIII (blue-green) layers, which emphasize gas content. Dust content is also emphasized, using archival infrared layers from NASA's Spitzer (8.0 micron, orange) and

These interfaces evolved a great deal in response to feedback from the educators, and the content of the modules that they developed (see Curriculum, above) evolved with them. Changes included adding a second HR diagram, at different wavelengths, to better constrain isochrone models (Cluster Two, Figure 1), cross-matching student-collected Skynet data with astronomy's Gaia catalog to remove field stars (Cluster Pro, Figure 2), adding multiple astronomical catalogs (Gaia, APASS, 2MASS, and WISE) to extend the interface's capabilities beyond what students can do with Skynet data alone (Cluster Pro Plus, about to be released), and adding a sonification feature to the Pulsar interface (Figures 8 – 10), which should benefit low- and regular-vision students alike.

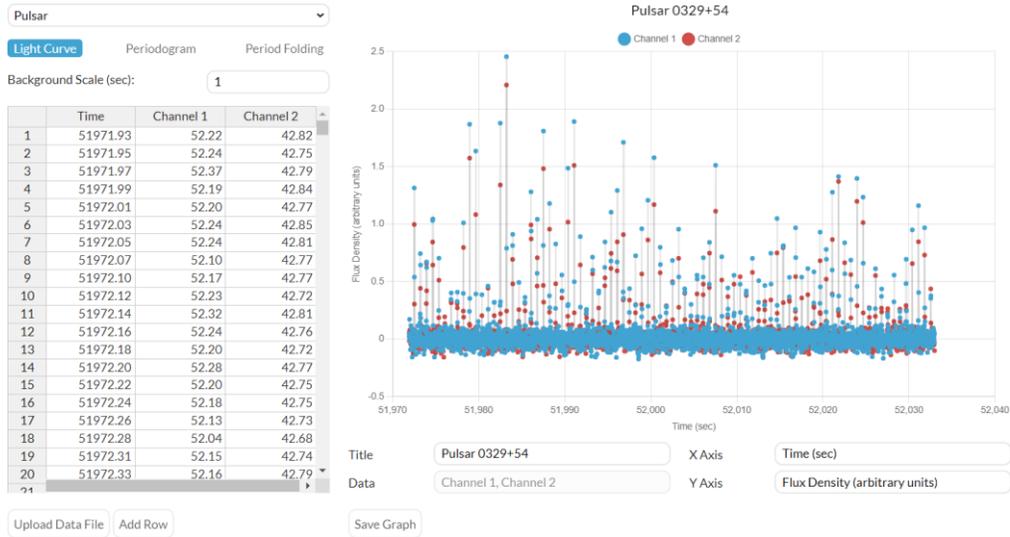

Figure 6. One minute of data from bright, slow pulsar 0329+54, collected with Skynet's 20-meter diameter telescope at Green Bank Observatory in West Virginia.

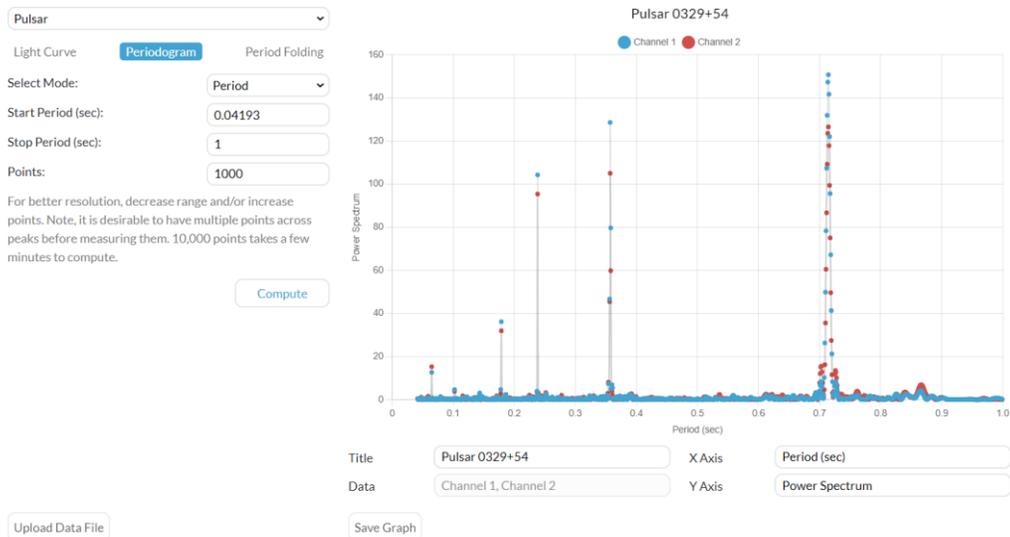

Figure 7. Periodogram of the data from Figure 6, computed using Lomb-Scargle (similar to a Fourier transform, but for non-uniformly sampled data). This pulsar is rotating every 0.7145 seconds.

While these tools were developed by undergraduate programmers, Skynet's two full-time, professional programmers pushed on other fronts.

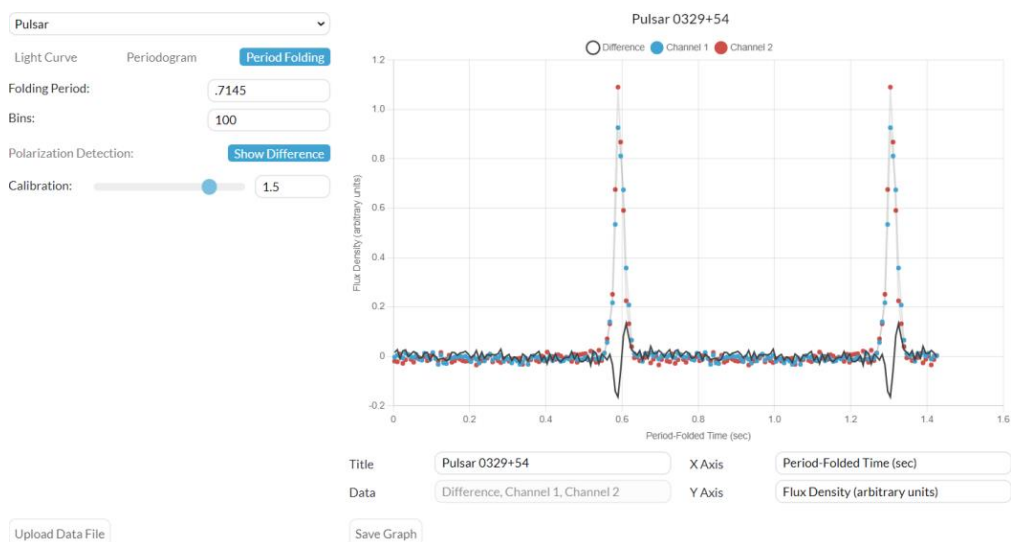

Figure 8. Period-folding of the data from Figure 7. The non-zero difference in the orthogonal polarizations channels shows that one channel's pulse leads the other's, demonstrating polarization and non-thermal emission (caused by electrons moving along the rotating neutron star's magnetic field lines).

First, they modified Afterglow Access (AgA) – our student-level, web-based, image processing and analysis application, originally developed with over $600K of funding from NSF – so students can not only carry out photometry, but calibrated photometry, of thousands of stars per image, for many images simultaneously. This was a significant effort, without which the Module 2 Cluster interfaces described above would be useless.

Students using AgA can now tie their photometric measurements to a wide selection of modern catalogs, allowing them to not only complete MWU! exercises, but to transition to publishable research using the exact same application. As part of this effort, we added professional-level photometry and source-extraction algorithms as well.

Much of our software effort in the second half of Y1 has focused on developing and adding astrophotography capabilities to AgA. These tools are needed in all of the modules, except for Module 4.

Of course, multiple astrophotography software packages already exist. However, they come with steep learning curves (and in places, favor less physically insightful approaches than we would prefer). Plus, our students will already be familiar with AgA from MWU!'s precursor curriculum, OPIS!, and will already be using AgA for other aspects of MWU!. As such, we have been developing and adding student-appropriate astrophotography capabilities to AgA. Our goal is for students to be able to achieve ≈85% of the quality of, e.g., something that might be accepted to NASA's Astronomy Picture of the Day, but with only ≈15% of the steps/complexity.

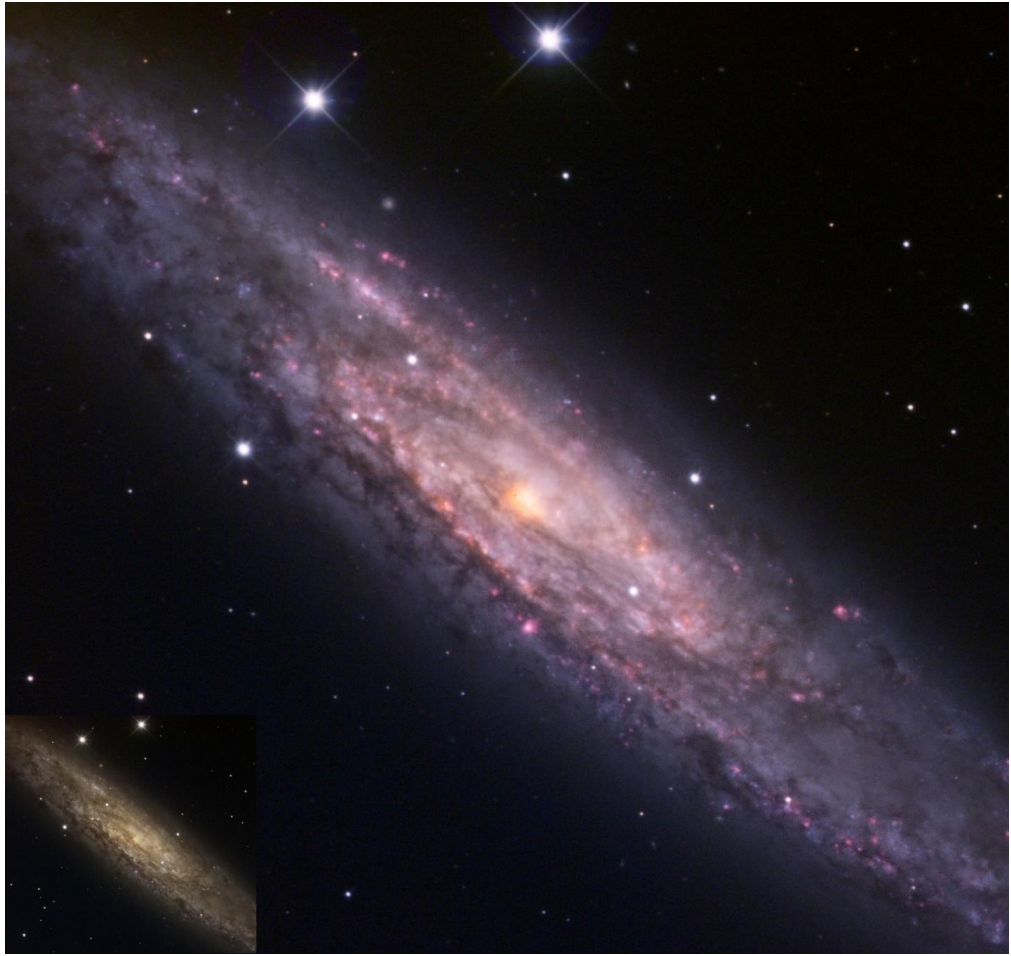

Figure 9. LRGB+NB+IR combination of NGC 253, using Skynet's PROMPT-6 telescope at CTIO, and supplemented with archival infrared data. All processing was done using AgA. Here, natural colors are achieved only after compensating for both dust within our galaxy (inset) *and* dust within NGC 253 (main image). SFRs are emphasized using the Hα layer (pink), and warm dust is emphasized using archival, infrared data from NASA's Spitzer spacecraft (8.0 micron, orange). The progression from dark, dusty clouds, to pink SFRs, to young, blue stars can be seen across each spiral arm (and understood in terms of spiral density waves).